\documentstyle[aps,multicol,psfig]{revtex}
\def\LamF{{\lambda_{\rm F}}}
\def\Kf{{k_{\rm F}}}
\def\Ef{{k_{\rm F}^{2}}}
\def\Hand{\hat{H}}

\def\radel{\rangle}
\def\thalp{\Theta}
\newcommand{\eqbreak}{
\end{multicols}
\widetext
\noindent
\rule{.48\linewidth}{.1mm}\rule{.1mm}{.1cm}
}
\newcommand{\eqresume}{
\noindent
\rule{.52\linewidth}{.0mm}\rule[-.1cm]{.1mm}{.1cm}\rule{.48\linewidth}{.1mm}
\begin{multicols}{2}
\narrowtext
}

\newcommand{\tc}{{T_{\rm c}}}
\newcommand{\tkt}{T_{\rm BKT}}
\newcommand{\xf}{\xi_{\phase}}
\newcommand{\phdag}{{\phantom{\dagger}}}

\newcommand{\kf}{k_{\rm F}}
\newcommand{\ef}{k_{\rm F}^{2}}

\newcommand{\curG}{{\cal G}}
\newcommand{\curO}{{\cal O}}

\newcommand{\tr}{{\rm Tr}}
\newcommand{\phase}{\vartheta}
\newcommand{\imag}{{\rm Im}}

\newcommand{\kb}{k_{\rm B}}

\newcommand{\brho}{{\bbox{\rho}}}
\newcommand{\BdG}{Bogoliubov--de~Gennes}
\input{epsf}
\begin{document}
\draft
\title{Probing d-wave pairing correlations in the pseudogap regime 
of the cuprate superconductors via low-energy states near impurities}
\author{Daniel E.~Sheehy\rlap,\cite{REF:DES}
\.{I}nan\c{c} Adagideli\rlap,\cite{REF:IA}
Paul M.~Goldbart\rlap,\cite{REF:PMG}
and
Ali Yazdani\cite{REF:AY}
}
\address{Department 
of Physics and Materials Research Laboratory, \\
University of Illinois at Urbana-Champaign,
Urbana, Illinois 61801, USA}
  \date{June 1, 2001}
\maketitle
\begin{abstract}
The issue of probing the pseudogap regime of the cuprate 
superconductors, specifically with regard to the existence and 
nature of superconducting pairing correlations of d-wave symmetry, 
is explored theoretically. 
It is shown that if the d-wave correlations believed to describe 
the superconducting state persist into the pseudogap regime, but 
with pair-potential phase-fluctuations that destroy their long-range 
nature, then the low-energy quasiparticle states observed near 
extended impurities in the truly superconducting state should also 
persist as resonances in the pseudogap regime.  
The scattering of quasiparticles by these phase-fluctuations broadens 
what was (in the superconducting state) a sharp peak in the 
single-particle spectral function at low energy, as we demonstrate 
within the context of a simple model.   
This peak and its broadening is, in principle, accessible via 
scanning tunneling spectroscopy near extended impurities in the 
pseudogap regime. 
If so, such experiments would provide a probe of the extent to which 
d-wave superconducting correlations persist upon entering the pseudogap 
regime, thus providing a stringent diagnostic of the phase-fluctuation 
scenario.
\end{abstract}
%
%
%
%
%
%
\begin{multicols}{2}
\narrowtext
\section{Introduction\/}
\label{SEC:intro}
Among the challenges presented by the high-temperature 
superconductors, one of the most persistent concerns the anomalous 
properties of the {\it normal\/} state of these materials.  In 
particular, the suppression of single-particle spectral weight around 
the Fermi energy~\cite{AGL96,HD96,CROF98} for temperatures above
the superconducting transition temperature $\tc$ of the underdoped 
cuprates indicates that the electronic behavior of these materials 
deviates substantially from that of conventional superconductors. 
There have been several theoretical scenarios proposed to explain 
this so-called pseudogap behavior~\cite{note:scenarios}.  In the 
present Paper, we shall be concerned with a particular one of these, 
viz., the phase-fluctuation scenario~\cite{EK95,FM98,KD98,REF:Loktev}.
According to this scenario, superconducting correlations in the 
form of Cooper pairing are presumed to exist (and to be responsible 
for the loss of single-particle spectral weight) for temperatures 
$T$ below the pseudogap onset temperature $T^{*}$.  However, in the 
intermediate temperature range (i.e.~$\tc< T\alt T^{*}$) the 
long-range spatial and temporal coherence in the phase of  
Cooper-pair wave functions (occurring for $T<\tc$) is presumed to  
present only up to intermediate length scales, having been disrupted 
on longer length scales
by the presence and motion of vortex excitations.  In other words, 
although long-range phase coherence is absent in it, the pseudogap 
regime is quantitatively distinguished from the conventional 
non-superconducting state by the presence of substantial, residual, 
local d-wave pairing correlations. 

Several recent experimental investigations support the notion of the 
phase-fluctuation scenario as the origin of  pseudogap phenomenology, 
including Refs.~\cite{REF:Corson,REF:Xu}.  Further steps towards an 
understanding the nature of the pseudogap regime would be furnished 
by experimental probes that are targeted towards the question of the 
existence of the putative local superconducting 
correlations~\cite{Janko,Choi,Sheehy,REF:Martin}.  Such probes would 
have the potential to discriminate between scenarios based on pairing 
correlations and those in which the pseudogap is due to some other 
mode of electronic ordering~\cite{note:scenarios}.  The purpose of 
the present Paper is to identify one such probe: scanning tunneling 
spectroscopy (STS) measurements of the single-particle spectral function 
near extended impurities in the pseudogap regime~\cite{eximp}. 

Before explaining the nature of this probe, let us pause to recall 
one of its essential ingredients.  It is has long been 
recognized~\cite{REF:Hu,REF:Sauls,REF:Fogel97}
that the scattering of quasiparticles between states corresponding to 
differing signs of the d-wave pair-potential leads to the existence 
of low-energy states~\cite{REF:BalatCollabs}.  Such scattering, and 
hence low-energy state-formation, occurs, e.g., at suitably oriented 
surfaces in the cuprates, leading to the observed 
zero-bias anomaly in the tunneling conductance~\cite{REF:Greene}.  In 
the setting of impurity (rather than boundary) scattering in d-wave 
superconductors, low-energy states, which in this case are localized 
near the impurities, have been observed in STS 
experiments~\cite{REF:Yazdani,REF:JCDavis}, 
and have been discussed theoretically for the case 
of point-like~\cite{REF:Balatsky} 
and extended~\cite{Ref:Inanc,Ref:SUSYnote} impurities.  
We remind the reader that these states co-exist with the continuum of 
low-energy quasiparticle states associated with the nodes at which the 
d-wave pair-potential vanishes.  Distinguishing between impurity states 
and nodal states is straightforward because the former give a peaked 
contribution to the spectral function whereas the contribution from the 
latter vanishes linearly at zero energy.  (A clear example of this is 
furnished by the data reported in Ref.~\cite{REF:Yazdani}.)

Returning to our main task, viz., probing the pseudogap regime for 
pairing correlations, we now state the central idea on which the 
present Paper is based.  Let us suppose that the pseudogap state is 
indeed distinguished by the presence of local (but not long range) 
d-wave pairing correlations.  The lack of long-range phase coherence 
in such a state complicates the experimental observation of the more 
conventional signatures of superconducting correlations (such as the 
Meissner effect).  However, the low-energy quasiparticle states 
occurring near extended impurities are {\it localized\/} in space, i.e., 
their existence only relies on the presence of local d-wave pairing 
correlations.  In consequence, these states should be only weakly 
affected by the destruction of long-range superconducting order that 
occurs at $\tc$, and hence should \lq\lq survive\rq\rq\ the transition 
into the pseudogap state.  Thus, if STS experiments were to reveal a 
sharp feature in the single-particle spectral function at low voltage 
bias, this would be evidence for the presence of local d-wave pairing 
correlations.  And if such experiments were able to characterize the 
temperature- and doping-dependence of the width of this spectral 
feature, this would provide a characterization of the nature of 
these finite-range spatial and temporal correlations.  
Of course, the interpretation of any experiment conducted at nonzero 
temperature would have to contend with broadening (arising, e.g., 
from thermal fluctuations of the sample).  In order to minimize the 
consequent smearing of the spectral function, which has the potential 
to obscure the very feature being sought, it would be preferable to 
examine cuprate systems having low values of $\tc$ (e.g.~heavily 
underdoped systems).  

Why are we focusing on the case of (non-magnetic) {\it extended\/} 
impurities as a probe of the pseudogap regime?  The main reason 
underlying this choice is as follows.  In the setting of extended 
impurities (as well as boundaries) in the superconducting state, it 
has been shown that the existence of low-energy quasiparticle states 
is a direct consequence of the d-wave nature (or, more precisely, the 
sign-changing nature) of the pair-potential~\cite{Ref:Inanc}.
Extending this reasoning to the case of extended impurities in the 
pseudogap regime, there the existence of low-energy states strongly 
depends on the presence of superconducting correlations of d-wave type. 

We note that Kruis et al.~\cite{REF:Kruis} have studied the density of 
states near a {\it point-like\/} impurity in the context of a simple 
phenomenological picture of the pseudogap regime.  In this picture, the 
physics of the pseudogap regime is incorporated through the hypothesis 
that, in the absence of the impurity, the single-particle density of 
states vanishes linearly at the Fermi energy.  At present, the extent 
to which an approach based on this picture can yield information about 
pairing correlations in the pseudogap regime is not clear.

The present Paper is organized as follows.  
In Sec.~\ref{SEC:Model} we provide a framework for discussing 
the influence of local d-wave pairing correlations on quasiparticle 
states near extended impurities~\cite{note:Extended}, focusing on the 
single-particle spectral function.  As we shall see, our expression 
for this spectral function will take the form of a density of states 
(determined at a fixed, locally phase-randomized d-wave pair-potential) 
averaged over the fluctuations of the phase field. 
In Sec.~\ref{SEC:semiclassical} we develop a semiclassical scheme 
for computing this density of states at fixed pair-potential in which 
we treat the long-wavelength pair-potential phase variations via 
perturbation theory.  This scheme allows us to focus on the 
contribution to the density of states at low energies, which involves 
states arising from changes in sign (as the momentum is varied) of 
the local pair-potential.  At this point we will have obtained an 
expression for the spectral function which consists of a sum of terms 
each associated with one classical scattering trajectory that passes 
through the vicinity of the impurity potential.   
In Sec.~\ref{SEC:gauss} we perform the average over the configurations 
of the fluctuating pair-potential arrived at via locally randomizations 
of the phase, under the assumption that the distribution of phase-field 
configurations is Gaussian.  We make an approximation to the resulting 
expression for the spectral function that is valid for the case of 
phase correlations persisting beyond the Cooper-pair size, finally 
arriving at an expression for the spectral function near an extended 
impurity that consists of a Gaussian peak the linewidth of which is 
proportional to the typical gradient of the phase of the pair-potential.  
In this section we also compute the linewidth of the spectral function 
in the pseudogap regime, doing so by assuming that the phase 
fluctuations are governed by the BKT theory of the two-dimensional 
$XY$-model.  finally, in Sec.~\ref{SEC:concluding} we make a numerical 
estimate of the linewidth by invoking the results of recent 
high-frequency conductivity data~\cite{REF:Corson}, and provide some 
concluding remarks.
\section{Model of cuprates with an extended impurity}
\label{SEC:Model}
In the present section we formulate the task of obtaining the 
single-particle spectral function in settings of systems of 
fermions interacting via some fermion-fermion coupling and 
also interacting with an external single-fermion potential 
(which could, e.g., represent an impurity potential). 
As our aim is to address the phase-fluctuation picture of the 
pseudogap regime of the cuprates, 
we envision following the standard field-theoretic route 
(see, e.g., Ref.~\cite{REF:Kleinert}) of exchanging the 
fermion-fermion coupling for a suitable collective 
quantum field $\Delta$.  Thus we arrive at the following 
formula for the one-fermion Green function 
$\curG(x,x')$:
\begin{mathletters}
\begin{eqnarray}
\label{eq:funcint}
&&
\curG(x,x')=\frac{
\displaystyle 
\int 
D\Delta^{\dagger}
D\Delta\,
{\rm e}^{-S[\Delta]}\,
G(x,x';\Delta)}
{\displaystyle
\int 
D\Delta^{\dagger}
D\Delta\,
{\rm e}^{-S[\Delta]}},
\\
\label{eq:gphi}
&&
\pmatrix{\partial_\tau+\hat{h} & \hat{\Delta} \cr
         \hat{\Delta}^{\dagger} & \partial_\tau-\hat{h}}
G(x,x';\Delta) = - \delta(x - x'),
\end{eqnarray}
\end{mathletters}
i.e., a \BdG\ Green function for pair-potential 
$\Delta$, suitably averaged over quantum field $\Delta$, the 
action for which is $S[\Delta]$.
Here, $x\equiv({\bf r},\tau)$ where ${\bf r}$ and $\tau$ are 
respectively the spatial position (in the two-dimensional 
${\rm CuO}_{2}$ plane) and the Matsubara time.
Furthermore, $\hat{h}\equiv-\nabla^{2}-\Ef+V({\bf r})$, 
in which $\Ef$ is the chemical potential
[i.e.~$\Kf$ ($\equiv 2\pi/\LamF$) is the Fermi wave vector],
$V$ is the single-particle potential, and we have adopted units
in which $\hbar^2/2m=1$,
$m$ being the (effective) mass of the electrons and holes.
The operator $\hat{\Delta}$ is the pair-potential (integral)
operator; how it acts is specified by the nonlocal kernel
$\Delta(x,x')$ via
$[\hat{\Delta}v](x)=\int dx'\Delta(x,x')\,v(x')$.

Our primary interest is in spectral function $\rho(E)$ in the 
pseudogap regime as well as the superconducting-to-pseudogap  
transition regime.  This quantity can be obtained in the 
usual way from $\curG(x,x')$ as follows: 
\begin{mathletters}
\begin{eqnarray}
\label{eq:DOS}
&&\rho(E) 
\equiv
-\pi^{-1} \imag\, \tr\int d^2 r \,
\curG({\bf r},{\bf r}; \omega_n)|_{i\omega_n \rightarrow E + i\delta} ,
\\
\label{eq:DOStwo}
&&
\curG({\bf r},{\bf r}'; \omega_n) 
\equiv
\int_0^{\beta}  
d\tau \,{\rm e}^{i\omega_n \tau} \curG({\bf r},{\bf r}';\tau,0),
\end{eqnarray}
\end{mathletters}
where $\tr$ denotes a trace only in the $2\times2$ particle-hole space,
$\beta\equiv 1/T$ (i.e., we have chosen units in which 
Boltzmann's constant $\kb =1$) and the $\omega_n\equiv (2n+1)\pi T$ 
(with $n$ integral)are fermionic Matsubara frequencies,and $\delta=0+$.  
We shall assume that the temperature is sufficiently high to validate 
the neglect of nonzero Matsubara-frequency modes of $\Delta$, which 
amounts to treating $\Delta$ as a classical (i.e.~non-quantal) 
statistical field.  Under this static condition, in which the fermion 
dynamics takes place in the presence of an unchanging $\Delta$ 
field, the spectral function $\rho(E)$ may be expressed in the form
\begin{mathletters}
\begin{eqnarray}
\rho(E)&=&
\langle\rho(E;\Delta)
\radel,
\label{eq:rhodef}
\\
\rho(E;\Delta)&=&
\sum\nolimits_n \delta(E-E_n),
\label{eq:rhoedelta}
\end{eqnarray}
\end{mathletters}
where $\{E_n\}$ is the collection of energy eigenvalues of the 
following \BdG\ eigenproblem in the presence 
of a generic classical configuration of $\Delta$:
\begin{equation}
\pmatrix{\hat{h} & \hat{\Delta} \cr
         \hat{\Delta}^{\dagger} & -\hat{h}}
\pmatrix{u\cr v}
=E\pmatrix{u\cr v}.
\label{EQ:1stOBdG}
\end{equation}
The notation $\langle\cdots\radel$ denotes the 
aforementioned static average over $\Delta$, i.e.,
\begin{equation}
\langle \curO  \radel \equiv
\frac{\int D\Delta^{\dagger}D\Delta\, {\rm e}^{-S[\Delta]}\, \curO}
     {\int D\Delta^{\dagger}D\Delta\, {\rm e}^{-S[\Delta]}}.
\end{equation}
(We do not specifically indicate that this functional 
average is only over static configurations of $\Delta$.)\thinspace\ 
Thus we have expressed the single-particle spectral function 
$\rho(E)$ in terms of a suitably averaged density of states 
$\rho(E;\Delta)$ for 
a corresponding \BdG\ eigenproblem. 

Before proceeding with the analysis of the eigenproblem given in 
Eq.~(\ref{EQ:1stOBdG}), we address the issue of the form of the 
pairing fluctuations that contribute dominantly to the average in 
Eq.~(\ref{eq:rhodef}).  This amounts to a statement about the 
physical physical picture of the pseudogap regime that we are 
concerned with.  Now, as discussed shortly before Eq.~(\ref{eq:rhodef}), 
we are considering only {\it static\/} configurations of $\Delta$,  
and therefore we shall henceforth simplify the notation by writing 
the pair-potential kernel $\Delta(x,x')$  as
$\Delta(x,x')=\Delta({\bf r},{\bf r}')$.  
In addition, it is convenient to transform  $\Delta({\bf r},{\bf r}')$ 
to relative and center-of-mass coordinates, $\brho$ and ${\bf R}$:
\begin{equation}
\bar{\Delta}(\brho,{\bf R})
\equiv
\Delta({\bf r},{\bf r}'),
\quad 
\brho \equiv{\bf r}-{\bf r}',
\quad
2{\bf R}\equiv{\bf r}+{\bf r}'\,. 
\end{equation}
It is then convenient to introduce the Fourier transform of 
$\bar{\Delta}$ with respect to the relative coordinate $\brho$, 
viz., 
$\bar{\Delta}({\bf k},{\bf R})$: 
\begin{equation}
\bar{\Delta}({\bf k},{\bf R})\equiv\int d^2\!\rho\,
{\rm e}^{-i{\bf k}\cdot \brho} \bar{\Delta}(\brho,{\bf R}), 
\label{eq:deltaft}
\end{equation}
thus obtaining a pair-potential at center-of-mass position ${\bf R}$ 
and relative momentum ${\bf k}$.  

As mentioned in Sec.~\ref{SEC:intro}, the scenario for the pseudogap 
regime with which we are concerned is based on the dominance of 
configurations of $\bar{\Delta}({\bf k},{\bf R})$ of the form 
\begin{equation}
\label{eq:phase}
\bar{\Delta}({\bf k},{\bf R})=
\bar{\Delta}_{0}({\bf k},{\bf R})\,
{\rm e}^{i\phase({\bf R})}.
\end{equation}
Here, the non-fluctuating factor 
$\bar{\Delta}_0({\bf k},{\bf R})$ is taken to have d-wave form 
(and can therefore be taken to be real)~\cite{note:rdep}. 
The fluctuating factor $\exp i\phase({\bf R})$ varies slowly with 
${\bf R}$.  By assuming this form for $\bar{\Delta}({\bf k},{\bf R})$
we are adopting the physical picture of the state as being one in 
which there is no long-range superconducting order, but there are 
local d-wave superconducting correlations, embodied in 
$\bar{\Delta}_0({\bf k},{\bf R})$~\cite{REF:swave}.  

In the following section, we obtain $\rho(E;\Delta)$ by making use of 
an elaboration of Andreev's semiclassical approach~\cite{Ref:Andreev} 
to the \BdG\ eigenproblem.   This elaboration is appropriate for the 
setting at hand, viz., one in which there is a strong single-particle 
potential.  This scheme was used in Ref.~\cite{Ref:Inanc} in order to 
address the low-energy density of states near an extended impurity in 
a d-wave superconducting state having a negligibly fluctuating, 
well-formed condensate.  
In contrast, our focus here is on situations in which the fluctuations 
in the {\it amplitude\/} of $\Delta$ are small, but the {\it phase\/} 
of $\Delta$ is strongly fluctuating.  Thus, although there is local 
pairing the system does does not exhibit long-range order.
\section{Semiclassical approach to the 
Bogoliubov--de~Gennes eigenproblem}
\label{SEC:semiclassical}
In Sec.~\ref{SEC:Model} we showed how the single-particle spectral 
function $\rho(E)$ can be expressed as a density of states $\rho(E;\Delta)$ 
for the \BdG\ eigenproblem at arbitrary pair-potential $\Delta$, averaged 
over $\Delta$ with a suitable weight, provided $\Delta$ can be treated 
in the static approximation.
To make progress with this \BdG\ eigenproblem we invoke, in the present 
section, a semiclassical approximation under which Eq.~(\ref{EQ:1stOBdG}) 
reduces to a family of one-dimensional eigenproblems labeled by the 
classical scattering trajectories of a particle in the presence of the 
single-particle potential $V({\bf r})$.
This approximation scheme, which was developed in Ref.~\cite{Ref:Inanc}, 
is valid provided that $\Ef\gg\{\Delta_{0},E\}$ for physically relevant 
configurations of $\Delta$ (where $\Delta_{0}$ is the magnitude of $\Delta$), 
and provided that $V({\bf r})$ and relevant configurations of $\Delta$ 
vary slowly on the scale of the Fermi wavelength $\lambda_{\rm F}$.

\begin{figure}[hbt]
\epsfxsize=\columnwidth
\centerline{\epsfbox{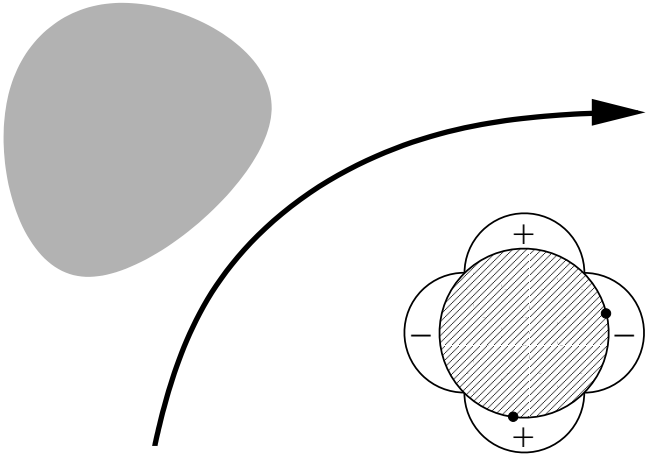}}
\vskip0.50cm
\caption{Schematic illustration of a classical trajectory for a 
quasiparticle scattering from an extended impurity.  Also shown 
is a depiction of the background d-wave pair-potential; 
dots indicate incoming and outgoing quasiparticle momenta.
\label{FIG:scatter}}
\end{figure}
We now turn to the family of one-dimensional eigenproblems arising
in our semiclassical scheme.  Following Ref.~\cite{Ref:Inanc}, it is 
straightforward to determine that the emerging trajectory-dependent
eigenproblem has the form
\begin{mathletters}
\begin{eqnarray}
\label{EQ:2ndOBdG}
&&\Hand
\pmatrix{\bar{u} \cr \bar{v}}
=
E\pmatrix{\bar{u} \cr \bar{v}},
\\
&&\Hand
\equiv
\pmatrix{-2i\Kf\,\partial_{s} &
\Delta_{0}(s)\exp{ i\phase(s)}\cr
\Delta_{0}(s)\exp{-i\phase(s)} &
2i\Kf\,\partial_{s}},
\\
&&\Delta_{0}(s)
\simeq
\bar{\Delta}_{0}
(\kf\partial_{s}{\bf x}_{\rm c}(s),{\bf x}_{\rm c}(s)).
\label{eq:deltaeff}
\end{eqnarray}
\end{mathletters}
Here, $\phase(s)\simeq\phase({\bf x}_{\rm c}(s))$, and
the parameter $s$ measures the position along a particular
classical trajectory ${\bf x}_{\rm c}(s)$, the latter obeying
the Newton equation 
\begin{equation}
\ef\,\partial_{s}^{2}{\bf x}_{\rm c}(s)=
-{\bbox{\nabla}}V({\bf x}_{\rm c}(s)).
\end{equation}
Each such classical trajectory (and associated eigenproblem) is labeled 
by an impact parameter $b$ and an incoming momentum direction ${\bf n}$. 

Our next task is to obtain the low-energy eigenvalues associated with 
Eq.~(\ref{EQ:2ndOBdG}) for the case of a d-wave pair-potential subject 
to generic spatial phase variations, i.e., Eq.~\ref{eq:phase}. 
To do this, we perform a local unitary transformation of $\Hand$, i.e., 
$\Hand \rightarrow  U\,\Hand\,U^{\dagger}$, 
where 
\begin{equation}
U(s) \equiv \pmatrix{ {\rm e}^{-i\phase(s)/2}& 
0\cr 
0& {\rm e}^{i\phase(s)/2 }},  
\end{equation}
and thus our eigenproblem acquires the form
\begin{mathletters}
\begin{eqnarray}
&&(\hat{H}_0 + \hat{H}_{1})
\pmatrix{
\bar{u}\cr
\bar{v}}=E
\pmatrix{
\bar{u}\cr
\bar{v}},
\label{EQ:3rdOBdG}
\\
&&\hat{H}_0(s)\equiv
\pmatrix{-2i\Kf\,\partial_{s}&
\Delta_0(s)\cr
\Delta_0(s)
&2i\Kf\,\partial_{s}},
\\
&&\hat{H}_{1}(s) \equiv
\pmatrix{\kf\,\partial_{s}\phase&
0\cr
0
&\kf\,\partial_{s}\phase}.
\end{eqnarray}
\end{mathletters}
The Hamiltonian for this eigenproblem now consists of a term 
arising from the underlying d-wave pair-potential (i.e.~$\hat{H}_0$) 
as well as a term that contains all the phase-variation information 
(i.e.~$\hat{H}_{1}$).  Our strategy is to treat $\hat{H}_{1}$ within 
perturbation theory, the starting point for which is the 
identification of the eigenstates of $\hat{H}_0$.  
As our purpose is to address low-energy states, it is adequate for us 
to focus on the {\it zero-energy\/} eigenstate of $\hat{H}_0$, if any 
there be.  As we shall discuss below, such states are guaranteed to 
arise for certain classical trajectories 
${\bf x}_{\rm c}(s)$~\cite{REF:Witten81,Ref:Inanc}.
The condition for the existence of such a state is determined by the 
asymptotic properties of $\Delta_0(s)$:  
If $\Delta_0(s)$ changes sign an odd number of times along an entire 
trajectory then the Hamiltonian $\hat{H}_0$ associated with that 
trajectory has precisely one zero-energy eigenstate.  Said more 
formally, if 
$\Delta_{\pm}\equiv\lim_{s\to \pm\infty}\Delta_0(s)$ then 
the condition for the existence of the zero-energy state is that
$\Delta_{+}\Delta_{-} < 0$.
The explicit form of this eigenstate is
\begin{mathletters}
\begin{eqnarray}
\label{eq:wavefunction}
\psi_{\pm}(s)  &=& \frac{1}{\sqrt{2}}
\pmatrix{ \phantom{\pm i}\varphi_{\pm}(s) \cr
          \pm i \varphi_{\pm}(s)},
\\
\label{eq:wavefunction2}
\varphi_{\pm}(s)
&\equiv& \alpha_{\pm} \exp  \pm(2 \kf)^{-1}
\int^{s}ds'\,\Delta_0(s')\, ,
\end{eqnarray}
\end{mathletters}
where $\psi_{+}$ corresponds to the case 
$\Delta_{+} < 0$ (so that $\Delta_{-} > 0$) and
$\psi_{-}$ corresponds to the case 
$\Delta_{+} > 0$ (so that$\Delta_{-} < 0$).
The pre-factors $\alpha_{\pm} $ are normalization factors, 
chosen so that
$\int_{-\infty}^{\infty} \varphi_{\pm}(s)^2 ds = 1$.  

Let us emphasize some of the important features of these
zero-energy states that hold provided the scatterer creates 
only trajectories with at most a single sign change in the 
pair-potential~\cite{note:moresignchanges}. 
First, the presence or absence of these states depends {\it only\/} 
on the properties of the d-wave pair-potential far from any impurity, 
and is insensitive to any amplitude variations of the d-wave 
pair-potential that might occur in the vicinity of this impurity.  
Second, the wavefunctions $\psi_{\pm}(s)$ exhibit exponential decay 
away from the impurity with a decay constant of order 
$\kf/|\Delta_{\pm}|$, which is proportional to the BCS correlation 
length.  As this length-scale is known to be short in the cuprates, 
these states are indeed an extremely {\it local\/} probe of local 
superconducting correlations.  


In the absence of phase fluctuations (i.e.~in the pure d-wave 
superconductor), these zero-energy eigenstates lead to a sharp peak 
in the density of states at low-energies~\cite{Ref:Inanc}. 
To access the impact of order-parameter phase variations on this peak, 
we include the effect of $\hat{H}_1$ on the zero-energy state
$\psi^{\dagger}_{\pm}(s)$ within quantum-mechanical perturbation theory.  
Thus, the shift in energy of the previously zero-energy eigenstate 
associated with a particular trajectory ${\bf x}_{\rm c}(s)$ 
(parametrized by ${\bf n}$ and $b$) is given by 
\begin{mathletters}
\begin{eqnarray}
\label{eq:perturbation}
\epsilon({\bf n},b)
&\equiv& \int\,ds\,
\psi^{\dagger}_{\pm}(s)\,
\hat{H}_{1}(s)\,
\psi^{\phdag}_{\pm}(s),
\\
&=&
\kf\int ds\,\partial_s\,\phase(s)\,\varphi_{\pm}(s)^2,
\label{eq:eps2}
\end{eqnarray}
\end{mathletters}
where we have used Eq.~(\ref{eq:wavefunction}) to get 
from Eq.~(\ref{eq:perturbation}) to Eq.~(\ref{eq:eps2}).
The spectral function at fixed pair-potential $\rho(E,\Delta)$ 
consists of contributions from all trajectories ${\bf x}_{\rm c}(s)$.  
As we are restricting our attention to low energies, 
it is sufficient to consider only trajectories for which 
$\Delta_{+}\Delta_{-}<0$ (i.e., those for which $\hat{H}_0$ 
has a zero-energy state).  The contribution to the low-energy 
spectral function at constant pair-potential due to the perturbed
zero-energy states is given by
\begin{eqnarray}
\label{eq:rhodelta}
&&\rho(E,\Delta)\simeq 
\kf \int \frac{d{\bf n}}{2\pi} \, \int db\,
\delta\big(E-\epsilon({\bf n},b)\big) 
\nonumber
\\
&&\hskip 3cm\times
\left(1-
{\rm sgn}\,\Delta_{+}\,\,
{\rm sgn}\,\Delta_{-}
\right),
\end{eqnarray}
where the factor
$\left(1-{\rm sgn}\,\Delta_{+}\,{\rm sgn}\,\Delta_{-}\right)$ 
ensures that only trajectories that satisfy $\Delta_{+}\Delta_{-}<0$   
(i.e.~exhibit an asymptotic sign change in $\Delta_0(s)$) 
contribute.  

It is important to emphasize that Eq.~(\ref{eq:rhodelta}) only 
includes the contribution from the perturbed zero-energy states, 
and that there will be additional, smaller contributions to the 
density of states at low energies.  Sources of these include 
remnants of the near-nodal quasiparticles states that exist in 
pure, d-wave superconductors at arbitrarily low energies. 
\section{Averaging over phase fluctuations}
\label{SEC:phasefluct}
In Sec.~\ref{SEC:semiclassical} we obtained an expression for the 
\BdG\ density of states $\rho(E,\Delta)$ in the presence of an 
extended impurity for the case of a pair-potential with local d-wave 
character and specific realization of the spatially varying phase 
$\phase({\bf r})$; see Eq.~(\ref{eq:rhodelta}).
In the present section, we calculate the spectral function $\rho(E)$ 
in the pseudogap regime by averaging $\rho(E,\Delta)$ over 
suitable configurations of $\phase({\bf r})$, i.e., by inserting 
Eq.~(\ref{eq:rhodelta}) into Eq.~(\ref{eq:rhodef}).  We remind the 
reader that our scheme for computing $\rho(E)$ applies to settings in 
which the temperature is high enough that we may treat the phase 
fluctuations classically, as discussed after Eq.~(\ref{eq:DOStwo}). 
\subsection{Gaussian model for phase fluctuations}
\label{SEC:gauss}
In order to evaluate the average over configurations of the 
pair-potential in Eq.~(\ref{eq:rhodef}) explicitly, we need a model 
for the weight of the various configurations.  For the sake of 
simplicity, we choose the phase-field configurations to have a 
(zero mean) Gaussian distribution characterized by the correlator 
$\langle
\nabla_{\alpha}\phase({\bf r})
\nabla_{\beta} \phase({\bf r}')
\radel$.
In terms of this correlator, it is straightforward to show that 
the spectral function $\rho(E)$ is a superposition of Gaussian 
distributions, one associated with each classical trajectory 
on which $\Delta_{0}$ changes sign: 
\begin{mathletters}
\begin{eqnarray}
\label{eq:DOSpseudogap2}
\rho(E)
&\simeq&
\kf \int \frac{d{\bf n}}{2\pi} \, \int db\,
\frac{\exp\left({-E^2/2\langle \epsilon({\bf n},b)^2\radel}\right)}
{\sqrt{2\pi\langle \epsilon^2 \radel}}
\\
&&\qquad\qquad
\times
\left(1-{\rm sgn}\,\Delta_{+}\,{\rm sgn}\,\Delta_{-}\right),
\nonumber 
\\
\langle \epsilon({\bf n},b)^2 \radel 
&\!\equiv\!&\kf^2 
\int\!\!ds\!
\int\!\!ds'\,
\varphi_{\pm}^2(s) \,
\varphi_{\pm}^2(s')\,
\langle
\dot{\phase}(s )\,
\dot{\phase}(s')
\radel.
\label{eq:width}
\end{eqnarray}
\end{mathletters}
Here, $\dot{\phase}(s)\equiv\partial_s \phase(s)$, and the width 
$\langle\epsilon({\bf n},b)^2 \radel$ of each Gaussian contribution 
is sensitive to the degree to which local superconducting 
correlations have been disrupted, as can be seen by the presence of 
the phase-phase correlator 
$\langle \dot{\phase}(s)\,\dot{\phase}(s')\radel$ 
in Eq.~(\ref{eq:width}). 
Our next task, then, is to evaluate the integrals over the 
parameters $s$ and $s'$ in Eq.~(\ref{eq:width}) in order to determine 
the width $\langle\epsilon({\bf n},b)^2 \radel$ associated with each 
trajectory $({\bf n},b)$.  In the next section, we carry out this 
evaluation within an approximation that is valid for the case of 
long-wavelength phase fluctuations.
\subsection{Approximate evaluation of trajectory integrals}
\label{SEC:trajectory}
In Sec.~\ref{SEC:gauss}, we obtained the 
expression~(\ref{eq:DOSpseudogap2}) for the spectral function 
$\rho(E)$ near an extended impurity in the pseudogap regime.  
In the present 
section, we make an approximation to our expression for $\rho(E)$ 
that makes use of the {\it local\/} nature of the low-energy states, 
as well as the long-wavelength nature of the pair-potential phase 
variations (appropriate for $T\agt \tc$).  
By {\it local\/} we mean 
that the wave functions $\varphi_{\pm}(s)$ exhibit exponential decay 
over a length-scale $\xi\sim\kf/|\Delta_{\pm}|$ (where $\Delta_0$ is 
the bulk d-wave pair-potential), i.e., the BCS correlation length; 
this can be seen by examining Eq.~(\ref{eq:wavefunction2}); 
see~\cite{REF:nodedir}.
In the cuprate superconductors, this length-scale is expected to be 
much shorter than the length-scale $\xf$ for typical pair-potential 
phase variations (i.e.~the inter-vortex spacing).  Thus one has a 
separation of length-scales: $\xf>\xi$.  
(Such a separation is a natural ingredient of the phase-fluctuation 
picture of the pseudogap regime because for inter-vortex spacings on 
the order of $\xi$ the conventional meaning of local Cooper pairs to 
break down.)\thinspace\   

To make the approximation to our expression for $\rho(E)$, consider 
the integrations over the trajectory parameters $s$ and $s'$ in 
Eq.~(\ref{eq:width}).  Now, the correlation function 
$\langle\dot{\phase}(s)\,\dot{\phase}(s')\radel$ varies appreciably 
only over length-scales on the order of $\xf$ or longer, whereas the 
wavefunctions $\varphi_{\pm}$ decay exponentially, as mentioned, 
on the length-scale $\xi$.  Hence, one can make an asymptotic 
approximation to the $s$ and $s'$ integrations which amounts to pulling 
the correlator out of the integrals.  Thus, owing to the normalization 
of $\varphi_{\pm}$, we have 
\begin{equation}
\langle\epsilon^{2}\radel 
\simeq
\kf^{2}\langle
\dot{\phase}(0)\,
\dot{\phase}(0)\radel,
\label{eq:epssquared}
\end{equation}
independent of ${\bf n}$ and $b$.   

Next, we turn to the interpretation of the quantity 
$\langle \dot{\phase}(0)\, \dot{\phase}(0) \radel$.
The derivative of the phase along a particular trajectory 
is given by 
\begin{equation}
\dot{\phase}(s)=
\dot{\bf x}_{\rm c}(s)\cdot 
\bbox{\nabla}
\phase({\bf x})
\big\vert_{{\bf x}={\bf x}_{\rm c}(s)},
\label{eq:phasederiv}
\end{equation}
By inserting Eq.~(\ref{eq:phasederiv}) into 
Eq.~(\ref{eq:epssquared}),
we see that the correlator of interest is 
$\langle \nabla_{\alpha}\phase({\bf r})
\nabla_{\beta}\phase({\bf r})\radel$ which, 
by spatial isotropy, has the form 
$\delta_{\alpha \beta}\, 
\langle\vert\bbox{\nabla}\phase({\bf r})\vert^2\radel/2$. 
According to the phase-fluctuation scenario, vortex excitations
are the dominant mechanism for generating phase gradients .  

How does this information about the phase correlations translate 
into information about the spectral function?  By noting that 
trajectories ${\bf x}_{\rm c}(s)$ involving sign-changes in the 
d-wave pair-potential obey $\vert\dot{\bf x}_{\rm c}\sim 1$ 
and using the aforementioned phase-gradient correlator, we 
arrive at the width
\begin{equation}
\langle\epsilon^{2}\radel
\sim
{\kf^2}\langle\vert
\bbox{\nabla}\phase({\bf r})
\vert^2\radel/2
\label{eq:widthfinal}
\end{equation}
and, hence, our final expression for the spectral function 
$\rho(E)$: 
\begin{eqnarray}
\label{eq:DOSpseudogap3}
\rho(E)
&\simeq&
\frac{\exp\left({-E^2/2\langle\epsilon^{2}\radel}\right)}
{\sqrt{2\pi\langle\epsilon^2 \radel}}
\\
&&\qquad\times
\kf\int\frac{d{\bf n}}{2\pi}\,\int db\,
\left(1-
{\rm sgn}\,\Delta_{+}\,
{\rm sgn}\,\Delta_{-}
\right).
\nonumber 
\end{eqnarray}
The first factor is a Gaussian in the energy $E$ with a 
linewidth $\langle\epsilon^{2}\radel^{1/2}$
associated with the r.m.s.~fluctuations in the phase gradient.
The second factor (i.e.~the integral over the impact parameter 
$b$ and the incoming momentum direction ${\bf n}$) determines 
the scale for $\rho(E)$, essentially by counting the number of 
sign-changing trajectories, and is expected to be only weakly 
temperature dependent.  In the next section, we compute the 
linewidth within a Berezinski\u\i-Kosterlitz-Thouless-like model 
of phase fluctuations near $\tc$.
\subsection{Linewidth near the superconducting phase boundary}
\label{SEC:linewidth}
To make further progress, we now attempt to calculate the spectral 
function linewidth due to the  phase fluctuations accompanying the 
destruction of superconducting order in the neighborhood of $\tc$.  
We shall do this by choosing a particular weight for the phase 
fluctuations, viz., that associated with the two-dimensional 
$XY$-model~\cite{Ber70,KT73,REF:ChaLub}.  Our analysis is reminiscent of 
that due to Franz and Millis~\cite{FM98} who addressed the bulk 
single-particle spectral function in the pseudogap regime.  In the 
present context, this analysis is provided solely for illustrative 
purposes, and is only meant only to provide a rough estimate
of the linewidth. 

Let us consider the $XY$-model action, 
\begin{equation}
\label{eq:xyaction}
S[\Delta]=\frac{K}{2} 
\int d^2 r \,
\vert\bbox{\nabla}\phase\vert^2,
\end{equation}
where $K\equiv\rho_s(T)/T$ in which $\rho_s(T)$ is the 
temperature-dependent superfluid density~\cite{REF:PreRanderia}.  
Note that we are not suggesting that the true critical fluctuations 
of the cuprate superconductors necessarily lie in the universality 
class of the two-dimensional $XY$-model, but simply that the 
intermediate length-scale fluctuations proposed as leading to pseudogap 
phenomena may adequately be modeled by Eq.~(\ref{eq:xyaction}).   
The form for 
$\langle\vert\bbox{\nabla}\phase({\bf r})\vert^2\radel$ for the 
$XY$-model may be calculated following the Debye-H\"uckel-type 
analysis of Halperin and Nelson~\cite{Ref:HN79}, giving
\begin{equation}
\label{eq:rgcorrelator}
\langle\vert\bbox{\nabla}\phase({\bf r})\vert^2\radel
\simeq\frac{2}{\pi \xf^2 K^*}\ln \Lambda \xf,
\end{equation}
in which $K^*$ is the short length-scale stiffness 
(obtained using the Kosterlitz renormalization-group equations), 
$\Lambda$ is a short-distance cutoff, and $\xf$ is a length-scale
characterizing the typical inter-vortex spacing.  The principal 
temperature dependence in Eq.~(\ref{eq:rgcorrelator}) arises 
via $\xf$: near $\tkt$ [i.e.~the transition temperature of the 
model~(\ref{eq:xyaction}), which is expected to lie not far below 
$\tc$] $\xf$ is proportional to 
$\exp\sqrt{\thalp/(T-\tkt)}$,
where $\thalp$ is a constant of order unity 
(which we do not try to calculate in detail).

\begin{figure}[hbt]
\epsfxsize=2.5in
\centerline{\epsfbox{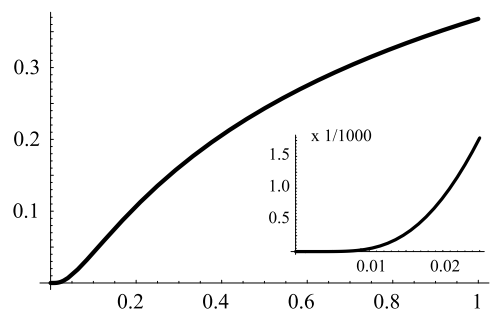}}
\vskip0.50cm
\caption{Spectral linewidth versus $(T-\tkt)/\thalp$.
Inset: Immediate vicinity of the transition.
\label{FIG:width_joint}}
\end{figure}
The proliferation of unbound vortex excitations upon warming 
through the BKT transition is reflected by this strong diminution 
of $\xf$ and causes a concomitant dramatic increase in the 
linewidth of the spectral function: 
\begin{equation}
\label{eq:enflucts}
\langle\epsilon^2 \radel^{1/2}
\propto
\exp\left(-\sqrt{\thalp/(T-\tkt)}\right). 
\end{equation}
Observation of a fluctuation-broadened peak in the spectral function, 
e.g., via STS measurements, would provide striking evidence in support 
of the phase-fluctuation scenario.  Moreover, the temperature dependence 
of the linewidth would, e.g., provide access to the details of the 
vortex-unbinding transition.
\section{Numerical estimate of the linewidth; Concluding remarks}
\label{SEC:concluding}
In the previous section we saw how one could estimate the 
temperature dependence of the linewidth of the spectral 
function near the superconducting transition temperature $\tc$.  
In the present section we make a rough numerical estimate of 
this linewidth at one particular temperature in the pseudogap 
regime by appealing to the data obtained in the high-frequency 
conductivity experiments of Corson et al.~\cite{REF:Corson} on 
${\rm Bi}_2{\rm Sr}_2{\rm Ca}{\rm Cu}_2{\rm O }_{8+\delta}$.  
We shall be specifically concerned with the interpretation of 
these data inasmuch as they provide access to the characteristic 
vortex density.

Until now we have been working with a system of units 
in which $\hbar^{2}/2m=1$.  Restoring conventional units in 
Eq.~(\ref{eq:widthfinal}) gives for the linewidth
\begin{equation}
\langle\epsilon^2\radel^{1/2}
\sim  
\frac{\hbar^2}{2m}\kf 
\langle\vert
\bbox{\nabla}\phase({\bf r})
\vert^{2}/2\radel^{1/2}.
\label{eq:widthfinal2}
\end{equation}
To estimate this width, we turn to the Corson et al.~experiments 
and its analysis by Corson et al., which is based on the notion 
that, at sufficiently high frequencies, the conductivity probes 
short-length-scale pairing correlations, and leads to an estimate 
for the characteristic density of free vortices $n_{\rm f}$.  Assuming 
that is is vortex excitations that lead to phase fluctuations, one 
expects that, up to a constant of order unity, 
$\langle |\bbox{\nabla}\phase({\bf r})|^2\radel\simeq n_{\rm f}$. 

Now, Corson et al.~obtain values of $n_f$ that are on the 
order of $0.003\,a_{\rm vc}$ (for $T\sim 75K$), 
where $a_{\rm vc}$ is the area of the 
core of a vortex.  
If we take the vortex core to be a disk of radius
$\xi\simeq 1\,{\rm nm}$, this leads to the value
$n_{\rm f}\sim 10^{3}\,\mu{\rm m}^{-2}$.
Then, using the order-of-magnitude estimate 
$\kf\simeq 1\,{\rm nm}^{-1}$, Eq.~(\ref{eq:widthfinal2}) gives 
$\langle \epsilon^2 \radel^{1/2}\simeq 9\,{\rm meV}$. 

The value of this estimate is that it shows that, for at least one 
cuprate material, there is a temperature at which the free-vortex 
density is small enough that phase fluctuations only weakly perturb 
the energies of the quasiparticle states.  Hence, the linewidth 
arising from perturbed zero-energy states can be rather smaller 
than the scale of the superconducting energy gap (and hence small 
enough to justify the our picture of perturbed zero-modes) but be 
large enough to be resolvable in STS measurements, such as those of 
Yazdani et al.~\cite{REF:Yazdani}.  Larger densities of free 
vortices, and hence large line-widths, would result from working at 
higher temperatures.  Moreover, near $\tc$ the vortex density is 
expected to show a very strong dependence on temperature, which 
should confer a strong temperature dependence on the linewidth.
Of course, if conducted at the temperature of the Corson et 
al.~experiments, thermal broadening would complicate the task of 
accessing the intrinsic linewidth (i.e.~the linewidth due to phase 
fluctuations); recall that $10\,{\rm K}$ is equivalent to 
$1\,{\rm meV}$.  
Thus, as emphasized in Sec.~\ref{SEC:intro}, one should consider 
performing experiments on materials having a lower $\tc$, so that 
the pseudogap regime can be explored at temperatures at which thermal 
broadening is less significant.  
These considerations indicate that it is at least conceivable that 
STS experiments near extended scatterers could provide a sharp test 
of the phase fluctuation scenario for the pseudogap regime.

\noindent
\section*{Acknowledgments}
\label{SEC:Acknowledgments}
It is a pleasure to thank Erich Mueller for discussions.  
This work was supported by the U.S.~Department of Energy, 
Division of Materials Sciences under Award No.~DEFG02-96ER45439,
through the Materials Research Laboratory at the 
University of Illinois at Urbana-Champaign 
(D.E.S., I.A., P.M.G., A.Y.), 
by NSF DMR99-75187 (P.M.G), and 
by NSF DMR98-75565 (A.Y.)
%
%

\end{multicols}

\begin{references}
\bibitem{note:scenarios}
Among these are SO(5)~\cite{REF:Zhang}, 
SU(2) slave boson~\cite{WL96}, 
pre-formed pairs~\cite{ML96},
antiferromagnetic spin fluctuations~\cite{SPS98},
and d-density wave~\cite{REF:Chakravarty}, 
to give a non-exhaustive list.
\bibitem{eximp}
By extended impurities we mean impurities
with a spatial extent that is large compared to the
Fermi wavelength $\lambda_{\rm F}$.
\bibitem{REF:BalatCollabs}
In the context of scattering due to point-like
impurities~\cite{REF:Balatsky},
the density of states possesses low-energy resonances, 
the energies of which move to zero energy as the 
strength of the impurity potential is increased.
\bibitem{Ref:SUSYnote}
The analysis of low-energy quasiparticle states reported in 
Ref.~\cite{Ref:Inanc} is based on an application of Witten's 
supersymmetric quantum mechanics~\cite{REF:Witten81}.
\bibitem{note:Extended}
Among the virtues of the setting of extended (rather than point-like) 
impurities are: 
(i)~the robustness of the low-energy states, by which we mean the 
insensitivity of their energies to spatial variations in the modulus 
of the d-wave pair-potential; and 
(ii)~the presence of low-energy states are signatures of 
superconductivity, specifically with sign changes in the 
pair-potential.
\bibitem{REF:nodedir} 
Of course, this exponential decay breaks down for trajectories in 
which at least one of the asympototic directions (incoming or outgoing) 
tends to a nodal direction.  Such trajectories comprise a relatively 
small fraction of the trajectories and will, henceforth, be neglected.
\end{references}
\end{document}